\documentclass[reprint, amsmath, amssymb, aps]{revtex4-2}
\usepackage{amsmath}
\usepackage{amsfonts}
\usepackage{graphicx}
\usepackage{appendix}
\usepackage{dsfont}
\usepackage{amssymb}
\usepackage{bm}
\usepackage{xcolor}
\usepackage{ulem}
\usepackage{soul}
\usepackage{mathrsfs}
\usepackage{amsmath}
\usepackage{esint}
\usepackage{natbib}
\usepackage{nccmath}
\usepackage{array}
\newcommand{\beq}{\begin{equation}}
\newcommand{\eneq}{\end{equation}}
\newcommand{\be}{\begin{equation}}
\newcommand{\ee}{\end{equation}}
\newcommand{\bea}{\begin{eqnarray}}
\newcommand{\eea}{\end{eqnarray}}

\usepackage{multirow}
\usepackage{wasysym}

\begin{document}

\title{Dissipation driven dynamical   topological phase transitions in 
two-dimensional superconductors}
 
\author{Andrea Nava$^{(1)}$,   Carmine Antonio Perroni$^{(2)}$,   Reinhold Egger$^{(1)}$, Luca Lepori$^{(3)}$, 
and Domenico Giuliano$^{(4)}$}
\affiliation{
$^{(1)}$Institut f\"ur Theoretische Physik IV,
Heinrich-Heine-Universit\"at, 40225 D\"usseldorf, Germany \\
$^{(2)}$ Dipartimento di Fisica ``E. Pancini'' Complesso Universitario Monte S. Angelo Via Cintia, I-80126 Napoli, 
Italy and \\ CNR-SPIN, Complesso Universitario Monte S. Angelo Via Cintia, I-80126 Napoli, Italy and  \\
I.N.F.N., Sezione di Napoli, Complesso Universitario Monte S. Angelo Via Cintia, I-80126 Napoli, Italy \\
$^{(3)}$Dipartimento di Scienze Matematiche Fisiche e Informatiche Universit\`a  di Parma and\\
  INFN, Gruppo Collegato di Parma, Parco Area delle Scienze 7/A, 43124, Parma, Italy. \\
$^{(4)}$Dipartimento di Fisica, Universit\`a della Calabria Arcavacata di 
Rende I-87036, Cosenza, Italy and \\ I.N.F.N., Gruppo collegato di Cosenza 
Arcavacata di Rende I-87036, Cosenza, Italy \\}

\begin{abstract}

We induce 
and study a topological dynamical phase transition between  two planar superconducting phases. Using the 
  Lindblad equation to account for the interactions of  Bogoliubov quasiparticles among themselves 
  and with the fluctuations of the superconducting order parameter, we derive  the relaxation dynamics 
  of the  order parameter. To  characterize the phase transition, we  compute the fidelity and the 
  spin-Hall conductance of the open system. 
  Our approach provides crucial informations  for experimental implementations, such as 
   the dependence of the critical time on the system-bath coupling.

\end{abstract}
\date{\today}
\maketitle

{\it Introduction.} Phase transitions (PTs) emerge as an effect of fluctuations, both thermal \cite{Itzykson1989} or quantum 
\cite{Sachdev2011}, involving collective degrees of freedom  in many-body systems. Typically, a PT is characterized by
the divergence of the correlation length as the temperature and/or another
control parameter is tuned from the outside and with the corresponding power-law scaling of the physical quantities, with 
``universal''  critical exponents. 
The usual approach to PTs in systems at equilibrium involves methods, such as looking for 
singularities and for critical scaling in the free-energy functional 
describing the system, as the control  parameters  are tuned close to their critical values.

A continuously increasing interest, both on the theoretical  as well as on 
the experimental side, has been recently gained by ``dynamical''  PTs in closed  many-particle 
quantum systems, prepared in a nonequlibrium state and then evolving in time with a pertinent
Hamiltonian \cite{Zvyagin2016,Heyl2018,Heyl2019}. In a DPT, it is the time $t$ that drives the system across criticality and the PT 
is evidenced by singularities in the matrix elements of the time evolution operator. To date, DPTs have been 
theoretically predicted and experimentally seen in various isolated quantum systems, in which   nonequilibrium
 is induced by quenching  some parameter(s) of the system  Hamiltonian 
\cite{Heyl2013,Jurcevic2017,Schmied2019,Yuzbashyan2006b,Mazza2017,Prufer2018,Yamamoto2021,Debashish2022,Debashish2023,Mazza2023}. 
 In a closed system, the DPT is analyzed by 
looking at the singularities in the  Loschmidt echo (LE) ${\cal L} (t) = | \langle \psi (0) |\psi (t) \rangle|^2$,
with $|\psi(t)\rangle$ being the state of the system at time $t$  and $|\psi (0)\rangle$ the pre-quench initial state
\cite{Zvyagin2016,Heyl2018,Heyl2019,Pollmann2010,Heyl2013}. While this 
approach has been also extended to the case in 
which the system has not been prepared in a pure state \cite{Abeling2016,Bhattacharya2017,Lang2018}, 
it does not apply to DPTs in open systems. The latter are  described by a time-dependent
density matrix $\rho (t)$, whose dynamics is determined by solving the pertinent evolution equation.

In the context of superconducting electronic systems, 
 nonequilibrium dynamics can be induced, for instance, by 
suddenly quenching the interaction parameter in a BCS Hamiltonian, and by encoding the 
following dynamical evolution of the system into an explicit dependence of the superconducting 
order parameter on $t$, by means of a time-dependent generalization of the self-consistent 
BCS mean-field (MF) approach \cite{Peronaci2015}. 

In this Letter we substantially extend and generalize the approach of  Ref.\cite{Peronaci2015},
so as to induce a DPT  between different superconducting
phases realized in a planar, interacting fermionic model with an attractive interaction.    
In particular, we   allow 
for the coexistence of $x^2-y^2$ ($d$-wave) and $xy$ ($id$-wave) superconducting gaps.
Then, after quenching the interaction strength(s) at time $t=0$, we  let the superconductor 
behave as an open system by 
exchanging Bogoliubov quasiparticles
 with the bath. In this way, we  
 account for the  dissipative dynamics induced by the interactions between
 quasiparticles not captured by the BCS approximation and/or by the coupling between the  fluctuations
 of the  order parameter and the quasiparticle continuum 
\cite{Yuzbashyan2005,Yuzbashyan2006,Yuzbashyan2006b,Cui2019}   and/or for the coupling to an 
external metallic contact \cite{Heimes2014} [a detailed discussion is provided in the Supplemental Material (SM), as well as
Refs.\cite{Yuzbashyan2005,Cui2019,Heimes2014,Nava2023}]. In particular, following Refs. \cite{Nava2021,Nava2023}, we 
do so within the Lindblad master equation (LME) approach to  the  time evolution of the superconductor 
density matrix $\rho (t)$. Our systematic approach naturally emerges from the microscopic model
of Ref.\cite{Heimes2014}. Moreover, it  is perfectly consistent with the one introduced in
 Ref.\cite{Cui2019} on phenomenological grounds, as discussed in Ref.\cite{longer_paper}, 
 where we also estimate typical experimental values of the coupling between the system and the bath. 
 
Here, we focus onto the DPT between a (topologically trivial)   
  $id$  and a  $d+id$  planar superconducting phase, the latter of which 
  is known  to describe the class C of topological planar superconductors,
  characterized by  particle-hole conjugation and  broken time-reversal symmetry 
    \cite{Schnyder2009,Bernevig2013,Goldman2016,Chern2016,Perroni2019,Lepori2021}. Therefore,   
    we realize a topological DPT (TDPT), to characterize which we first of all look at  the transition in time  
of the superconducting order parameter, between the  asymptotic values (at $t=0$ and 
$t \to\infty$), respectively corresponding to the $id$ and to the $d+id$ phase.
Then, rather than the LE, we approach the DPT   by 
using   the fidelity ${\cal F}(t)$ between the initial pure state and the one described by $\rho (t)$, which 
is more suitable for an open system  \cite{longer_paper}.
Finally, to make a rigorous statement on the topological properties of the phases separated by the 
DPT, we compute   
the spin-Hall conductance of the system as a function of $t$, $\sigma(t)$. In a stationary state,
$\sigma(t)$  is  proportional to the 
topological invariant whose nonzero value is a signal of a nontrivial topological phase \cite{Qi2010,Chern2016}.  
Furthermore, it is   well defined  even when the system goes through the DPT  \cite{Qi2010}.  
In addition to defining a protocol to monitor a DPT in an open system, 
a topic that has  recently become of the utmost relevance \cite{Wu2022}, our approach provides remarkable results
of practical interest in a possible experimental realization of the system that we study (e.g., in ultracold atom
lattices), such as the variation of the ``critical time'' $t_*$ as a function of the system-bath coupling. 

\begin{figure}
 \center
\includegraphics*[width=0.9 \linewidth]{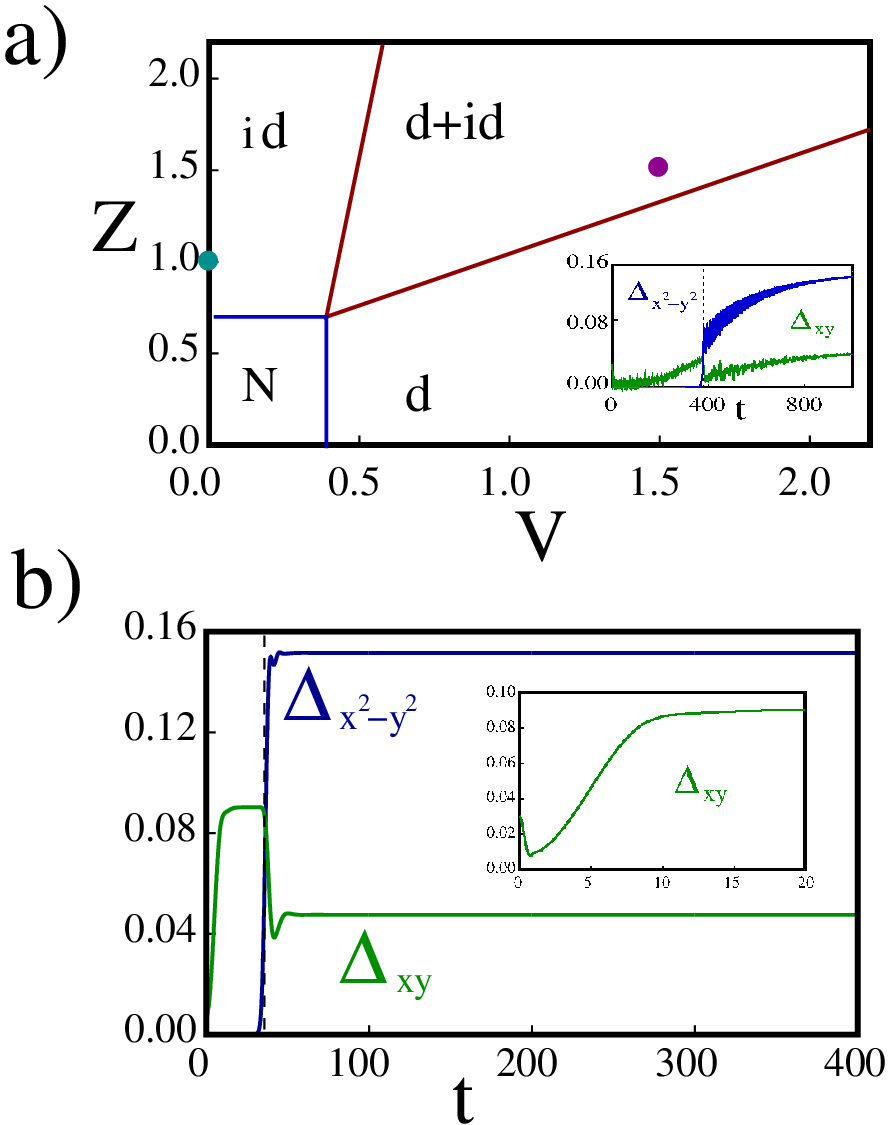}
\caption{{\bf a):}  Equilibrium phase diagram 
  in the $V-Z$ plane. The cyan (magenta) dot corresponds to $(V^{(0)},Z^{(0)})$  
    $(V^{(1)},Z^{(1)})$ (see text).   [{\bf Inset:} Same  as in panel {\bf b)} but with $g=0.002$]. 
{\bf b):} Time dependent gap $\Delta_{x^2-y^2} (t)$ (blue curve) and  $\Delta_{xy} (t)$ (green curve), for
 $V^{(1)}=Z^{(1)}=1.5$,    $g=0.2$,
 and prepared, at $t=0$, in a state with   $\Delta_{xy}^{(0)} \approx 0.03$. 
 [{\bf Inset:} Zoom of the   plot of $\Delta_{xy}(t)$ for
$0\leq t\leq20$].   } 
\label{phdiag}
\end{figure}

{\it Model Hamiltonian.} Our main reference Hamiltonian $H_{\rm MF}$
stems from the self-consistent mean-field (SCMF) approximation of the
 Hamiltonian describing interacting spinful fermions on a two-dimensional (2D) square
lattice introduced in Ref.\cite{longer_paper}, with 
a  nearest-neighbor  and  a  next-to-nearest neighbor density-density interaction, 
both attractive in the spin singlet channel, respectively, with interaction strengths  
    $V$ and $Z$ (both $>0$). We therefore set \cite{Peronaci2015,longer_paper}

\beq
H_{\rm MF}=\sum_{{\bf k}}[c_{{\bf k},\uparrow}^\dagger , c_{{\bf -k},\downarrow}  ]
\left[ \begin{array}{cc}  \xi_{\bf k} & - \Delta_{\bf k} \\ - [ \Delta_{\bf k}]^* & - \xi_{\bf k} \end{array} \right]
\left[ \begin{array}{c} c_{{\bf k},\uparrow} \\ c_{{\bf - k},\downarrow}^\dagger \end{array} \right], 
\label{eh.2abis}
\eneq
\noindent
with   ${\bf k}$ summed  over the full Brillouin zone and with  $c_{{\bf k},\sigma}$ and $c_{{\bf k},\sigma}^\dagger$
being  the fermion operators in ${\bf k}$
space.  The ${\bf k}$-dependent gap in 
 Eq.(\ref{eh.2abis}) takes, in general, a nonzero imaginary part. The two parameters  take the form 
\begin{eqnarray}
\xi_{{\bf k}}&=&-2 [ \cos(k_x)+\cos(k_y)]-\mu \; , \label{eh.4a}  \\
\Delta_{\bf k}&=&\ 2\Delta_{x^2-y^2} \{\cos(k_x)-\cos(k_y)\}-4i\Delta_{xy}\sin(k_x)\sin(k_y) 
\; ,
\nonumber
\end{eqnarray}
\noindent
with the chemical potential $\mu=0$  (half filling) and   $\Delta_{x^2-y^2}$ and $\Delta_{xy}$ determined by the SCMF equations \cite{longer_paper}

\begin{eqnarray}
\Delta_{x^2-y^2} &=& \frac{V}{4 {\cal N} } \sum_{\bf k}  \frac{ [\cos (k_x ) - \cos (k_y ) ] \Delta_{\bf k}}{\epsilon_{\bf k}}  \; ,  \nonumber \\
\Delta_{xy} &=& \frac{iZ}{2 {\cal N} } \sum_{\bf k}  \frac{ \sin (k_x ) \sin (k_y )  \Delta_{\bf k}}{\epsilon_{\bf k}}  
\;  , 
\label{eh.bis1}
\end{eqnarray}
\noindent
$ {\cal N}$ being the number of lattice sites, $\epsilon_{\bf k} = \sqrt{ \xi_{\bf k}^2 + | \Delta_{\bf k}|^2}$ and the 
lattice constant =1. Our model calculation encompasses all the relevant features
that should characterize a TDPT,  both in solid-state \cite{Heimes2014}, as well as in quantum-optical open systems 
\cite{Wu2022,Cardano2017,Derrico2020}.  In Fig.\ref{phdiag}{\bf a)} we plot the 
 phase diagram of our system in terms of 
  $V$ and $Z$ (which we regard as our physically tunable parameters) 
as determined by Eqs.(\ref{eh.bis1}). For small, though finite, 
values of  $V$ and $Z$ the system lies within a normal phase (N).
Keeping  $Z$  ($V$) small and increasing $V$ ($Z$), our system undergoes a phase transition, with a 
gap $\Delta_{x^2-y^2}$ ($\Delta_{xy}$) continuously developing a nonzero value for $V>V_c$ ($Z>Z_c$), with 
$V_c\approx 0.35$ ($Z_c\approx 0.7$), and with the gaps  increasing with $V$ and $Z$, according to 
Eqs.(\ref{eh.bis1}). At large enough values of both $V$ and $Z$, the system undergoes a topological phase
transition, at which a   $d+id$ phase opens, where both $\Delta_{x^2-y^2}$ and $\Delta_{xy}$ are $\neq 0$. The latter phase   
 exhibits nontrivial topological properties. 
We now show how to realize a  TDPT between the  $id$  and the 
 $d+id$  phase, along the dissipative dynamics of the nonequilibrium superconductor. 
 
{\it Dynamical phase transition:} To induce a DPT in our system,   we prepare it in  the ground state of 
 $H_{\rm MF}$  with  $\Delta_{x^2-y^2}^{(0)} =0$  
and $\Delta_{xy}^{(0)}\neq0$. At time  $t=0$, we quench the interaction strengths to
$ (V^{(1)}, Z^{(1)} )$, corresponding to  both $\Delta_{x^2-y^2}^{(1)} $ 
and $\Delta_{xy}^{(1)}$ being $\neq 0$. 
The  induced nonequilibrium dynamics  makes  $\Delta_{\bf k}$ to explicitly depend on $t$.
To  describe the time evolution of the open system, we extend the time-dependent 
SCMF approach of Ref.\cite{Peronaci2015}  by allowing 
 the system to exchange Bogoliubov quasiparticles with an external bath,
 thus  resorting to the LME approach to the density matrix dynamics (see Sec. III of SM).
 Following   Refs.\cite{Yuzbashyan2005,Yuzbashyan2006,Yuzbashyan2006b,Cui2019,Nava2021,Nava2023}, we 
 therefore write the LME for $\rho (t)$ as

\begin{eqnarray}
&& \frac{d \rho (t)}{dt} = -i[H_{\rm MF} (t),\rho(t)]+ g\sum_{\lambda = \pm}\sum_{\bf k} \label{leq.1} \\
&&\times [  f(-\lambda \epsilon_{\bf k} (t) ) (
 2 \Gamma_{{\bf k},\lambda} (t)  \rho (t) \Gamma_{{\bf k},\lambda}^\dagger (t)  -
\{  \Gamma_{{\bf k},\lambda}^\dagger (t)  \Gamma_{{\bf k},\lambda}  (t),\rho (t)  \}  ) 
  \nonumber \\
&&+ f(\lambda \epsilon_{\bf k} (t) ) ( 2 \Gamma_{{\bf k},\lambda}^\dagger (t) \rho (t) 
\Gamma_{{\bf k},\lambda}(t) - \{ \Gamma_{{\bf k},\lambda} (t) \Gamma_{{\bf k},\lambda}^\dagger(t) , \rho (t)\}  )]\nonumber 
\:  .
\end{eqnarray}
\noindent
In Eq.(\ref{leq.1})  $g$ is the strength of the system-bath coupling and, 
consistently with   the detailed balance principle, recovering 
 the Boltzmann distribution as a stationary solution of the LME is assured by our setting of  
the coupling strength corresponding to the Bogoliubov quasiparticle annihilation and creation 
operators, $\Gamma_{{\bf k},\lambda}$ and $\Gamma_{{\bf k},\lambda}^{\dagger}$,   
to be proportional to $f(-\lambda \epsilon_{\bf k})$ and to
$f(\lambda \epsilon_{\bf k})$, respectively,  with $f(\epsilon )$
  being the Fermi distribution function (note that we employ particle-hole symmetry 
  to  set $1-f(\epsilon)=f(-\epsilon)$) \cite{Petruccione2002,Nava2019}.
The time dependence of    $\Delta_{\bf k}(t)$  makes $H_{\rm MF} (t)$ in Eq.(\ref{leq.1}) 
as well as its eigenvalues ($\pm \epsilon_{\bf k}(t)=\pm \sqrt{\xi_{\bf k}^2 + |\Delta_{\bf k}(t)|^2}$) and  
 eigenmodes ($\Gamma_{{\bf k},\pm }(t)$),
 to acquire an explicit time dependence, as well (see SM and Refs.\cite{Peronaci2015,Nava2023} 
 for further details). 
 
 To complete the SCMF approach, we need  the  relation between $\Delta_{\bf k}(t)$ and 
  $\rho(t)$. To recover it,  we follow the derivation 
of Ref.\cite{Peronaci2015} by generalizing Eqs.(\ref{eh.bis1}) to self-consistent
relations between $\Delta_{\bf k} (t)$ and 
 $f_{\bf k} (t) ={\rm Tr} [\rho (t)  c_{{\bf -k},\downarrow}c_{{\bf k},\uparrow}]$, as 
 we discuss in detail in the SM.  
 
  In Fig.\ref{phdiag}{\bf b)} we plot $\Delta_{x^2-y^2} (t)$ and $\Delta_{xy} (t)$ in a system prepared in 
  the ground state $|\psi (0)\rangle$ of $H_{\rm MF}$ with   $(V^{(0)},Z^{(0)})=(0,1.0)$, corresponding to 
 $\Delta_{x^2-y^2}^{(0)}=0,\Delta_{xy}^{(0)}=0.03$.  At $t>0$ we quench  the interaction 
 strengths to $(V^{(1)},Z^{(1)})=(1.5,1.5)$, corresponding to  $\Delta_{x^2-y^2}^{(1)}=0.15,\Delta_{xy}^{(1)}=0.05$, 
  and let the system evolve according to  Eqs.(\ref{leq.1}), with 
 $g=0.2$ (main figure)  and $g=0.002$   (inset  of Fig.\ref{phdiag}{\bf a)}).
   Given $(V^{(0)},Z^{(0)})$ and  $(V^{(1)},Z^{(1)})$,  
    the time evolution of the superconducting gaps is directly determined by 
 Eqs.(\ref{leq.1}) and by the time-dependent generalizations of Eqs.(\ref{eh.bis1}) (see Eq.(\ref{s.0.4}) of SM). 
We see    that there is a  finite interval of time $[0,t_*]$, with $t_* \approx 40$,  within which 
$\Delta_{xy} (t)$ stays finite, and basically constant, while $\Delta_{x^2-y^2}(t)$ 
remains pinned at 0.  As $t$ goes across $t_*$ (vertical, dashed line), 
$\Delta_{xy} (t)$ almost suddenly lowers its
value, while $\Delta_{x^2-y^2} (t)$   switches from zero to a finite value, which 
keeps roughly constant for any $t>t_*$.   $t_*$   is determined by
the finite time required for the $f_{\bf k} (t)$'s   to take the appropriate 
threshold value to trigger the onset of the two-component order parameter.  
The sharp change in $\Delta_{\bf k} (t)$ across $t=t_*$ corresponds to a transition, in real time,
of the system between  two different phases characterized by different values of the superconducting 
order parameter, that is, to a DPT  \cite{Zvyagin2016,Heyl2018,Heyl2019}. The  relatively high
 value of $g$ (although still quite smaller  than any other energy scale in the system) induces a sharp switch 
in the values of the two different symmetry components of the order parameter at the DPT. 
To highlight the effects of varying $g$, in the inset  of Fig.\ref{phdiag}{\bf a)} we show the same plots
as in the main figure, but with $g=0.002$. In this case, $t_*$ becomes much larger than before and  
 $\Delta_{xy} (t)$ oscillates   and monotonically increases, starting  from
  $\Delta_{xy}^{(0)}$, as long as $t<t_*$. At the same time, $\Delta_{x^2-y^2}(t)=0$. 
At $t \geq t_*$, both $\Delta_{xy} (t)$ and $\Delta_{x^2-y^2}(t)$
undergo a discontinuous jump, after which they  
 start to oscillate around the values they  take in the asymptotic 
 ($t\to\infty$) state. Again, we  conclude that,  at $t=t_*$, our system goes across  a DPT,  
triggered by the mismatch between the initial  and the asymptotic state of the system.

\begin{figure}
 \center
\includegraphics*[width=.95 \linewidth]{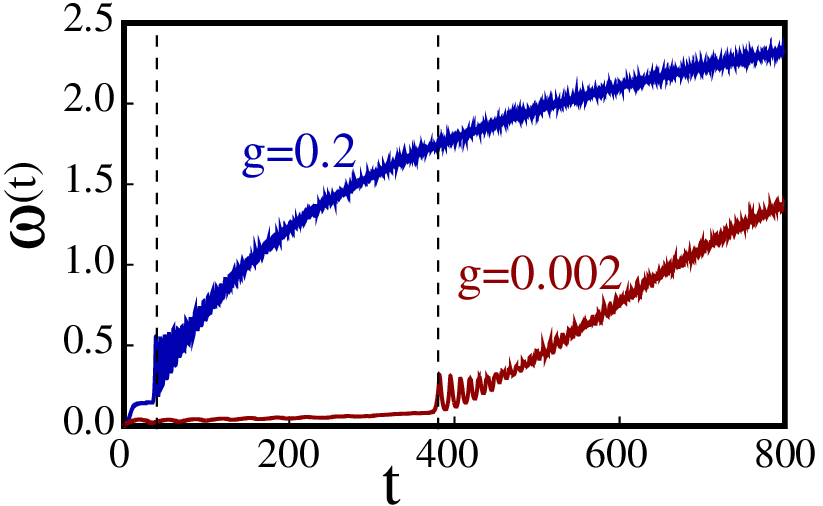}
\caption{$\omega (t)$ as a function of time $t$ computed with the time-dependent 
MF Hamiltonian with parameters $\Delta_{x^2-y^2} (t)$ and $\Delta_{xy} (t)$, as in 
Fig.\ref{phdiag}{\bf b)}, for $g=0.2$ (blue curve) and $g=0.002$ (red curve). 
  The dashed vertical lines mark the DPT.}
\label{figs_2}
\end{figure}

We evidence   the onset of the  DPT by looking 
at nonanalyticities in the fidelity ${\cal F} (t)$ 
between the initial state $ | \psi (0)\rangle$ and the state
described by $\rho (t)$, ${\cal F} (t) = \langle \psi ( 0) | \rho (t) | \psi ( 0 ) \rangle$. 
 Indeed, as pointed out in  Refs.\cite{Mera2018,Zhao2009}, at a DPT ${\cal F} (t)$ is expected to 
show nonanalyticities similar to what one would obtain in the LE, computed in a closed system. In fact, while 
the LE at time $t$ is defined if the system lies in a pure state at any $t$, when the 
system  state is described in terms of a  density matrix $\rho(t)$,  ${\cal F} (t)$ shows to be 
the pertinent quantity to evidence the DPT. 
In particular,  we look for nonanalyticities
 in the  rate function $\omega (t) = - \frac{1}{{\cal N}} \log [ {\cal F} (t)]$ 
\cite{Jurcevic2017,Zvyagin2016,Heyl2018,Heyl2019} (see \cite{longer_paper} for the 
  mathematical derivation ). In Fig.\ref{figs_2} we plot $\omega (t)$ as a function of $t$ in 
both cases corresponding to the plots in Fig.\ref{phdiag}. Aside from 
the  different scale $t_*$ at different values of $g$, we note that, 
 for $0 \leq t < t_*$, $\omega (t)$ takes a mild dependence on $t$, 
 with $\omega (t) \sim 0.1-0.2$, denoting an 
appreciable overlap between $|\psi ( 0)\rangle$ and the state at time $t$.  This
basically evidences the persistence of the system 
within the same phase 
\cite{Zanardi2006_2,Zanardi2006,Zvyagin2016,Heyl2018,Heyl2019}. At $t=t_*$, the sudden change in the slope of 
  $\omega (t)$  demonstrates how $t=t_*$ corresponds to a  nonanalyticity tied to the
DPT.  The subsequent rapid increase in $\omega (t)$ for $t>t_*$  corresponds
to a drastic reduction in ${\cal F} (t)$ (by orders of magnitude), which is a clear signal that, moving across $t=t_*$, the system 
has gone through a DPT.

\noindent

{\it Topological phase transition:} To physically
ground the topological nature of the DPT,  we now review the 
calculation of  the spin-Hall conductance
 $\sigma (t)$   across the DPT of our system (for details, see SM as well as Refs.\cite{Cardano2017,Derrico2020,Qi2010,Schnyder2009,Chern2016,Goldman2016,Kantian2009,Mazza2023}). 
    In an equilibrium  state  $\sigma (t)$ 
 is proportional to the Chern number ${\cal C}$: in the trivial phase, 
 ${\cal C}=0$, while in a topologically phase,   ${\cal C} = \pm 2$  \cite{Schnyder2009,Bernevig2013}. 
 Apparently, ${\cal C}$  is ill defined across 
 the DPT. Instead, $\sigma (t)$  is perfectly well defined 
and can be computed at any finite $t$ within linear (in the applied voltage bias) response theory. 
Following the  derivation of  the SM, 
 we note that Fig.\ref{phdiag}{\bf b)}  suggests that,  for $g=0.2$,  $\Delta_{x^2-y^2} (t)$ and  
 $\Delta_{xy} (t)$ can be well approximated 
as  $\Delta_{x^2-y^2}(t)=\theta (t-t_*) \Delta_{x^2-y^2}^{(1)}$, and 
$\Delta_{xy}(t)=\theta (t_*-t) \Delta_{xy}^{(0)}+\theta(t-t_*)\Delta_{xy}^{(1)}$, with $\theta(t)$ being Heaviside's step function, 
$\Delta_{x^2-y^2}^{(0)}=0$, $\Delta_{xy}^{(0)} =0.03$, $\Delta_{x^2-y^2}^{(1)}=0.15$, and $\Delta_{xy}^{(1)}=0.05$.  
 In Fig.\ref{total_sigma}  we plot  $\sigma (t)$ computed accordingly. 
As expected, for  $t<t_*$ $\sigma (t)=0$. Passing across the DPT at $t=t_*$, $\sigma(t)$ jumps to a finite value,
and then, for $t>t_*$,  it shows  damped  oscillations ($\sim (t-t_*)^{-1}$) toward  the asymptotic value
$\sigma_\infty = \lim_{t\to\infty}\sigma (t) =  2(2\pi)^{-1}$. This is exactly what is expected for a TDPT. In addition, we have also verified 
that $\sigma_\infty$ does not change on varying $\Delta_{x^2-y^2}^{(1)}$ and $\Delta_{xy}^{(1)}$, provided we stay within 
the $d+id$ phase in Fig.\ref{phdiag}{\bf a)}.

    \begin{figure}
 \center
\includegraphics*[width=.95 \linewidth]{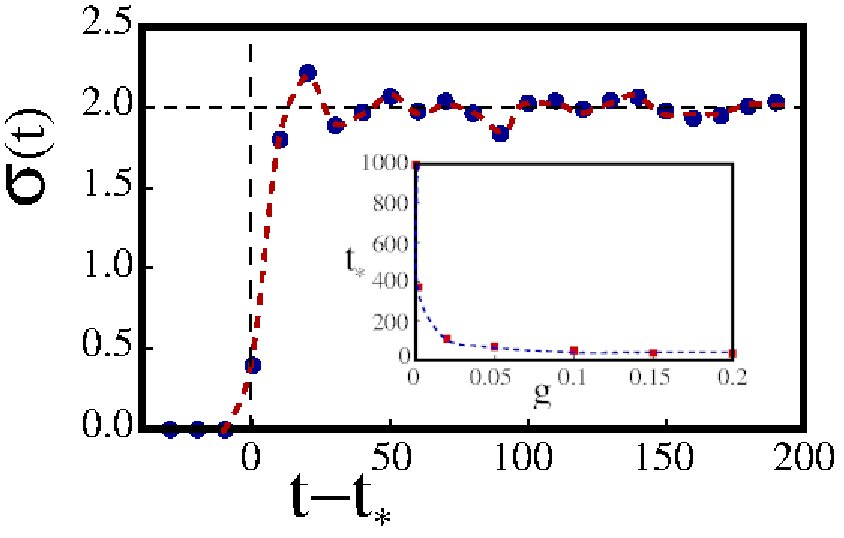}
\caption{ $\sigma (t)$ (in units of $(2\pi)^{-1}$) 
computed   
 in the sudden jump approximation (see SM), 
with
$\Delta_{xy}^{(0)} = 0.09$, $\Delta_{x^2-y^2}^{(1)}=0.15$,  $\Delta_{xy}^{(1)}=0.05$, and $g=0.2$.  
The dashed vertical line marks the DPT at $t=t_*$, the dashed horizonal line marks the  value of $\sigma_\infty$.
 [{\bf Inset}: $t_*$ computed in the same system for different values of $g$ (red squares)
 between $g=0.0005$ and $g=0.2$ (the dashed  lines are a guide to the eye).]}  
\label{total_sigma}
\end{figure}

While our sudden  jump approximation only applies for $g$ as large as 0.2,  from the inset of 
Fig.\ref{phdiag}{\bf a)} we see that, even for $g=0.002$, the main features of 
Fig.\ref{phdiag}{\bf b)} still persist, that is,  a sharp reduction in $\Delta_{xy} (t)$ and   a corresponding jump in $\Delta_{x^2-y^2} (t)$  from 
0 to a finite value at $t=t_*$.   From
the qualitative point of view, we can still get some hints by  enforcing the  approximation of  $\Delta_{\bf k}(t)$ with 
a piecewise function, although along small time intervals (depending, of course, on the frequency of the superimposed
oscillations). This would allow us 
to map out the full time dependence of $\sigma (t)$ on $t$. Eventually, we expect that 
$\sigma (t)$ will asymptotically converge to the value dictated by the asymptotic value of $\Delta_{x^2-y^2} (t)$ and
of $\Delta_{xy}(t)$ as $t\to\infty$.  Again, these correspond to 
a  $d+id$ superconducting state and, therefore, we find that  $\sigma (t) \to_{t \to \infty} 2(2\pi)^{-1}$, even at
values of $g$ smaller than 0.2 by  two orders of magnitude. 

{\it Conclusions:}  In this Letter we have shown that a DPT can take place in an open nonequilibrium planar superconducting system,
described by a  LME  that is determined by the tunnel coupling to an external metallic lead \cite{Heimes2014}, or mimics the 
residual quasiparticle interaction beyond BCS theory    \cite{Cui2019}.  
To monitor the system across the  DPT, we synoptically looked at the self-consistently computed 
 superconducting gap, at the fidelity,  and at the spin-Hall conductance.  In particular, this last quantity 
 is crucial in evidencing the topological nature of the DPT, as it crosses over from being zero for $t<t_*$  to 
 an asymptotic value $\sigma_\infty$, as $t\to\infty$, which corresponds to a topologically nontrivial $d+id$ phase.

Within our derivation,  we explicitly show for the first time a clear evidence that 
a  DPT can take place in an open  solid-state system, in particular between two phases with different topological properties. In doing so,
we also highlight the importance of  the system-bath coupling  in stabilizing a DPT transition 
between two selected phases with given properties. We do so by means of   a combined use of the LME and
 of the time-dependent SCMF approach, thus defining  a systematic framework to discuss 
 the peculiar properties of a DPT in an open system. 
Due to our  minimal set of assumptions, we believe 
that the range of applicability  of our approach is   much wider than just the one that
 we discuss here.  For instance,  it would be important to study whether 
 the topological DPT   takes place at any  finite $g$, or 
whether there is a critical value $g_c$ such that, for $g<g_c$, it is 
washed out by the uncontrolled oscillations in the superconducting order parameter  that arise at small values of 
$g$ \cite{Peronaci2015}.   Along this direction, in the inset of Fig.\ref{total_sigma} we report 
the results of a preliminary calculation of $t_*$ at selected 
values of $g$ between $g=0.0005$ and $g=0.2$. We check that a power-law fit of the dependence of 
$t_*$ on $g$ such as $t_*= Ag^{-B}$, with $A\approx 7.2$ and $B\approx 0.65$ fits the data reasonably well.
In addition, the strong increase of $t_*$ as $g\to 0$ is consistent with the absence of any DPT in the closed 
quenched superconductor studied in Ref.\cite{Peronaci2015}. However, providing a certain answer on those issues requires developing
a model calculation of $t_*$ at a given $g$, which is an interesting topic for future research. 

{\color{black}
\section*{Supplemental Material}
\label{SM}

In the following Supplemental Material we provide the details of our calculations that, although 
being technically important to recover our results, are not essential to follow our main derivation. In particular, 
in Subsection \ref{smsec.1} we discuss the main steps leading to our time-dependent self-consistent mean field approximation, 
in Subsection \ref{smsec.2} we present our derivation of the time-dependent spin-Hall conductance, in Subsection \ref{smsec.3} we 
derive the Lindblad Master Equation from a specific microscopic model of a two-dimensional superconductor coupled to a metallic 
lead and also discuss how the same equation can arise from the quasiparticle dynamics in the superconductor beyond BCS approximation, 
in Subsection \ref{smsec.4} we show our results for the rate function and for the spin-Hall 
conductance across a dynamical phase transition from 
a planar d+id to a planar s-wave superconducting phase.

  \subsection{Time-dependent self-consistent mean field approximation}
\label{smsec.1}

In order to summarize the main steps behind   Eq.(\ref{leq.1}), here we review the time-dependent 
SCMF approximation, as introduced 
in Ref.\cite{Peronaci2015} for an open, superconducting system, together with our
specific implementation in the context of the LME, which we derive in detail in Ref.\cite{longer_paper}.  

In order to induce the nonequilibrium dynamics of our system, we prepare it in some initial state, corresponding to 
the groundstate of a superconducting Hamiltonian such as the one in Eq.(\ref{eh.2abis}) with given interaction strengths 
 and then, at $t=0$, we quench the interaction strengths themselves. At the same time, 
we couple the system to the external bath. As a result, on one hand, the following time evolution of the  density matrix of the 
resulting open system is described by the LME  in Eq.(\ref{leq.1}), on the other hand, the superconducting gap parameter entering 
Eq.(\ref{leq.1}), $\Delta_{\bf k} (t)$, acquires an explicit dependence on $t$, which has to be 
self-consistently determined, in response to the quench in the interaction strengths. 

As discussed in Ref.\cite{longer_paper}, to  set up the time-dependent SCMF framework we assume that Eq.(\ref{leq.1}) holds 
and define $\nu_{\bf k}(t)$ and $f_{\bf k}(t)$ as 

\begin{eqnarray}
\nu_{\bf k}(t)&=&{\rm Tr} \left[\rho(t) \left( c_{{\bf k},\uparrow}^\dagger c_{{\bf k},\uparrow}-\frac{1}{2}\right)\right] \nonumber \\
f_{\bf k}(t)&=& {\rm Tr} \left[\rho(t)c_{{\bf -k},\downarrow} c_{{\bf k},\uparrow}\right] 
\;\;\; . 
\label{s.0.1}
\end{eqnarray}
\noindent
From Eq.(\ref{leq.1}), we derive the closed set of differential equations for $\nu_{\bf k} (t)$ and $f_{\bf k}(t)$ given by 

\begin{eqnarray}
\frac{ d \nu_{{\bf k} } (t)}{dt}&=& -\frac{g \xi_{\bf k}}{\epsilon_{\bf k}(t)} -2 g\nu_{\bf k}(t) + \Im m\{ [    \Delta_{\bf k} (t) ]^*   f_{\bf k} ( t) \}  
\label{s.0.2} \\
\frac{ d f_{\bf k} (t)}{dt}&=& -(2i \xi_{\bf k} +2g) f_{\bf k}(t) - 2 i \Delta_{\bf k} (t) \nu_{\bf k}(t) + \frac{g \Delta_{\bf k} (t)}{\epsilon_{\bf k}(t)} 
  \:\nonumber ,
\end{eqnarray}
\noindent
with $\epsilon_{\bf k}(t)=\sqrt{\xi_{\bf k}^2 + |\Delta_{\bf k}(t)|^2}$ and $\Im m$ denoting the imaginary part.  For a superconductor at 
equilibrium, $\Delta_{\bf k}$, $\epsilon_{\bf k}$ and $f_{\bf k}$, all independent of time, are related to each other by the condition 
$f_{\bf k}=\Delta_{\bf k}/\epsilon_{\bf k}$. In the nonequilibrium case, following Refs.\cite{Peronaci2015,longer_paper},  
we assume that  $\Delta_{\bf k}(t)$ takes an explicit dependence on $t$ according to  

\begin{eqnarray}
\Delta_{\bf k}(t)&=&2\Delta_{x^2-y^2} (t) \{\cos(k_x)-\cos(k_y)\} \nonumber \\
&-&4i\Delta_{xy}(t)\sin (k_x)\sin(k_y)
\;\;\; , 
\label{s.0.3}
\end{eqnarray}
\noindent
with  $\Delta_{x^2-y^2}(t)$ and $\Delta_{xy}(t)$ determined by  generalizing Eqs.(\ref{eh.bis1}), to account for
the explicit dependence on $t$, as  

\begin{eqnarray}
\Delta_{x^2-y^2}(t)&=&\frac{V}{4{\cal N}}\sum_{{\bf k}} [\cos(k_x)-\cos(k_y)]f_{{\bf k}}(t) \nonumber \\
\Delta_{xy}(t)&=&\frac{iZ}{2{\cal N}}\sum_{{\bf k}} \sin(k_x)\sin(k_y)f_{{\bf k}}(t) 
\:\:\: ,
\label{s.0.4}
\end{eqnarray}
\noindent
with ${\cal N}$ being the number of lattice sites. 
Having determined $\Delta_{\bf k}(t)$ from Eqs.(\ref{s.0.3},\ref{s.0.4}), we insert the corresponding result in 
$H_{\rm MF}$ in Eq.(\ref{eh.2abis}), thus defining the time-dependent mean-field Hamiltonian $H_{\rm MF}(t)$. 
Therefore, regarding $t$ as over-all parameter, we determine the eigenvalues and the corresponding 
quasiparticle eigenmodes of $H_{\rm MF}(t)$, respectively given by $E_{{\bf k},\pm }(t)=\pm\sqrt{\xi_{\bf k}^2+|\Delta_{\bf k}(t)|^2}$ and 
$\Gamma_{{\bf k},\pm} (t)$, both carrying an explicit dependence on $t$. In particular, the operators 
$\Gamma_{{\bf k},\lambda}(t)$ are recovered by means of a pertinent, time-dependent generalization of the Bogoliubov-Valatin 
transformations discussed in detail in Ref.\cite{longer_paper}. Inserting the time-dependent parameters and eigenmode operators
in the LME for the density matrix $\rho (t)$, we recover Eq.(\ref{leq.1}). Of course, to preserve self-consistency at any time $t$, 
Eq.(\ref{leq.1}) has to be solved together with the self-consistent Eqs.(\ref{s.0.4}). Due to the apparent complexity of the whole procedure, 
we resorted to a numerical approach to the solution of the coupled equations, to recover the results for the time-dependent 
quantities that we discuss in the main paper.

 \subsection{Derivation of the time-dependent spin-Hall conductance $\sigma(t)$}
 \label{smsec.2}

To discuss our derivation of  the spin-Hall conductance $\sigma (t)$, we refer to
 Fig.\ref{phdiag}{\bf b)}, where we plot $\Delta_{x^2-y^2}(t)$
and $\Delta_{xy}(t)$ for $g=0.2$. From the figure, we note that the time evolution of $\Delta_{\bf k}(t)$
can be regarded as made out of two regions ($0\leq t<t_*$ and $t_*<t$), in which both $\Delta_{x^2-y^2}$ and 
$\Delta_{xy}$ are constant, separated by a sudden   change  in the superconducting gaps at $t=t_*$. 
Therefore, we conclude that, at least for $g=0.2$,  
our approach explicitly realizes, in the case of a two-dimensional superconductor, 
the protocol outlined in Refs.\cite{Cardano2017,Derrico2020}, in a chiral quantum walk of twisted 
photons. In this specific case,   taking into account the sudden change in the parameters of 
 $H_{\rm MF}$ that takes place at $t=t_*$,  
 we describe the time evolution of our system by means of the 
time-dependent MF Hamiltonian $\hat{H}_{\rm MF} (t)$, given by 

\beq
\hat{H}_{\rm MF}  (t) = \sum_{\bf k} [c_{{\bf k},\uparrow}^\dagger,c_{ {\bf - k},\downarrow}]
\left[ \begin{array}{cc} \xi_{\bf k} & - \Delta_{\bf k} (t) \\ -
[\Delta_{\bf k} (t)]^* &-\xi_{\bf k}\end{array} \right]\left[\begin{array}{c}
c_{{\bf k},\uparrow}\\c_{{\bf -k},\downarrow}^\dagger \end{array}\right] 
\:,
\label{cch.1}
\eneq
\noindent
with   $\Delta_{\bf k}(t)=\theta (t_*-t)\Delta_{\bf k}^{(0)} + \theta (t-t_*)\Delta_{\bf k}^{(1)}$. 
 Of course, resorting to the sudden change approximation for $\Delta_{\bf k}(t)$ remarkably 
simplifies our analytical calculations of the spin-Hall conductance,  without  substantially affecting the 
final result for $\sigma(t)$. In fact, we find only minimal discrepancy between the spin Hall conductance 
computed with this approximation compared to the fully self-consistent result. 

  From Eq.(\ref{cch.1}) we obtain that the spin current operators now explicitly depend on $t$ as well. They
are given by ($a=x,y,z$)

\begin{eqnarray}
&& j_{{\rm sp},a} (t)  = \sum_{\bf k}  [c_{{\bf k},\uparrow}^\dagger,c_{{\bf -k},\downarrow}] \nonumber \\
&& \times \frac{\partial}{\partial k_a} \left[ \begin{array}{cc} \xi_{\bf k} & - \Delta_{\bf k} (t) \\ -
[\Delta_{\bf k} (t)]^* &-\xi_{\bf k}\end{array} \right]\left[\begin{array}{c}
c_{{\bf k},\uparrow}\\c_{{\bf -k},\downarrow}^\dagger \end{array}\right]   \:.
\label{cch.z3}
\end{eqnarray}
\noindent 
 The change in $\hat{H}_{\rm MF}$ at $t=t_*$ comes along with a similar change in the expressions 
for its eigenmodes, which we respectively 
 refer to as $\Gamma_{{\bf k},\lambda}^{(0)}$ and as $\Gamma_{{\bf k},\lambda}^{(1)}$. Specifically, we obtain

\begin{eqnarray}
\Gamma_{{\bf k},+}^{(u)} &=& \cos \left(\frac{\theta_{\bf k}^{(u)}}{2}\right)c_{{\bf k},\uparrow} - 
\sin \left(\frac{\theta^{(u)}_{\bf k}}{2}\right)
e^{i\phi^{(u)}_{\bf k}}c_{{\bf -k},\downarrow}^\dagger 
\label{cch.z4} \\
\Gamma_{{\bf k},-}^{(u)} &=& \sin \left(\frac{\theta_{\bf k}^{(u)}}{2}\right)
e^{-  i\phi^{(u)}_{\bf k}}c_{{\bf k},\uparrow} 
+ \cos \left(\frac{\theta^{(u)}_{\bf k}}{2}\right) c_{{\bf -k},\downarrow}^\dagger  \nonumber 
\;\;, 
\end{eqnarray}
\noindent 
with  $u=0,1$, and 

\begin{eqnarray}
 \cos (\theta_{\bf k}^{(u)})  &=& \frac{\xi_{\bf k}}{\epsilon_{\bf k}^{(u)}} \; , \nonumber \\
 \sin (\theta_{\bf k}^{(u)} ) e^{- i\phi_{\bf k}^{(u)}}  &=& - 
 \frac{\Delta_{\bf k}^{(u)}}{\epsilon_{\bf k}^{(u)}} \; , \nonumber \\
 \epsilon_{\bf k}^{(u)}&=& \sqrt{\xi_{\bf k}^2+|\Delta_{\bf k}^{(u)}|^2} 
 \:\:\:\: . 
 \label{cch.z4a}
 \end{eqnarray}
To describe the time evolution of the system coupled to the bath, in the following we use   a ``mixed'' representation, 
in which we let the current operators  $j^M_{{\rm sp},a}(t)$  evolve with $\hat{H}_{\rm MF} (t)$, and the density  matrix 
$\bar{\rho} (t)$ depend on time as an effect of the coupling to the  bath only.  This implies that, for 
$t>t_*$ and in the zero-temperature limit, $\bar{\rho} (t)$  satisfies the differential equation  

 \begin{eqnarray}
&& \frac{d \bar{\rho} (t)}{ d t } = 
 g \sum_{\bf k} \{[ \Gamma_{{\bf k},+}^{(1)},\bar{\rho}
 (t)[\Gamma_{{\bf k},+}^{(1)}]^\dagger]
-[[\Gamma_{{\bf k},+}^{(1)}]^\dagger,\Gamma_{{\bf k},+}^{(1)} \bar{\rho}(t)]\} \nonumber \\
&&+ g  \sum_{\bf k}  \{[[\Gamma_{{\bf k},-}^{(1)}]^\dagger ,\bar{\rho}(t) 
\Gamma_{{\bf k},-}^{(1)}]-[\Gamma_{{\bf k},-}^{(1)},[\Gamma_{{\bf k},-}^{(1)}]^\dagger \bar{\rho} (t)]\} 
\:\:\:\: ,
\label{cch.15}
\end{eqnarray}
\noindent
 which we will employ throughout this section to derive the time evolution of our system, 
consistently with the approximate expression for the time-dependent mean field
Hamiltonian, $\hat{H}_{\rm MF} (t)$, which we are using here. 
On the other hand, for $t<t_*$, we find $\frac{d \bar{\rho} (t)}{dt} = 0$ and hence $\bar{\rho} (t) = 
\bar{\rho} ( 0 ) = | \psi (0) \rangle \langle \psi (0) | $, with 
$\Gamma_{{\bf k},+}^{(0)} | \psi (0) \rangle = [\Gamma_{{\bf k},-}^{(0)}]^\dagger  | \psi (0) \rangle=0$ $\forall {\bf k}$. 
From Eq.(\ref{cch.15}) one readily derives 
 the time evolution of the covariance matrix element in the energy 
basis, $\bar{c}_{{\bf k};(\alpha , \beta )} (t)  = {\rm Tr} \{ [\Gamma_{{\bf k},\alpha}^{(1)}]^\dagger 
\Gamma_{{\bf k},\beta}^{(1)} \bar{\rho} ( t ) \}$ in the form 

\begin{eqnarray}
&& \bar{c}_{{\bf k};(\alpha,\beta)} (t) =  \bar{c}_{{\bf k};(\alpha,\beta)} ( 0) \theta (t_*-t) +  \theta(t-t_*) \times 
\label{axx.1}  \\
&&\{e^{-2g(t-t_*)} \bar{c}_{{\bf k};(\alpha,\beta)} ( 0) + [1-e^{-2g(t-t_*)}]\delta_{\alpha,-}\delta_{\beta,-} 
\} \nonumber
 \:\: . 
\end{eqnarray}

We now  compute $\sigma (t)$ by resorting to the standard    linear response theory
of charge transport. In particular, we apply a uniform external field in the $y$ direction at frequency $\omega_0$, 
compute the corresponding current response in the $x$ direction, and eventually recover $\sigma (t)$, which is 
given by 
 
\begin{eqnarray}
   \sigma  (t) = &-& i \lim_{\omega_0 \to 0 } \: \int_{-\infty}^t \: d \tau \: {\rm Tr} \{ [ j^{M}_{{\rm sp} , x} (t) , 
   j^{M}_{{\rm sp} , y} ( \tau ) ] \bar{\rho} (t ) \} 
  \nonumber \\
&\times& \frac{ \sin ( \omega_0 ( \tau-t_*) )}{ \omega_0} e^{ \eta (\tau - t_*)}  
\:\:\:\: , 
\label{cch.9}
\end{eqnarray}
\noindent
with  $\eta = 0^+$. The spin current operators $j^{M}_{{\rm sp},a} (t)$ taken in the mixed representation, 
as discussed above, are given by 

\begin{widetext}
\beq
 j^{M}_{{\rm sp},a}(t) = \theta (t_*-t) \sum_{{\bf k}} \sum_{\lambda , \lambda'} 
[j_{{\rm sp},a}^{(0)} ({\bf k} )]_{\lambda , \lambda'} [\Gamma_{{\bf k},\lambda}^{(0)}]^\dagger \Gamma_{{\bf k},\lambda'}^{(0)} 
e^{i\epsilon_{\bf k}^{(0)} (\lambda - \lambda')t} 
+ \theta (t-t_*) \sum_{{\bf k}} \sum_{\lambda , \lambda'} 
[j_{{\rm sp},a}^{(1)} ({\bf k} )]_{\lambda , \lambda'} [\Gamma_{{\bf k},\lambda}^{(1)}]^\dagger \Gamma_{{\bf k},\lambda'}^{(1)}
e^{i\epsilon_{\bf k}^{(1)} (\lambda - \lambda')t}
 \:\:\:,
\label{cxx.1}
\eneq
\noindent
\end{widetext}
with $[j_{{\rm sp},a}^{(0)} ({\bf k} )]_{\lambda , \lambda'}$ and $[j_{{\rm sp},a}^{(1)} ({\bf k} )]_{\lambda , \lambda'}$ 
denoting the matrix elements of the spin current  operators    for 
$t<t_*$ and for $t>t_*$, respectively, in the basis of the eigenmodes of $\hat{H}_{\rm MF}(t)$.

As long as $t<t_*$, the explicit calculation of $\sigma (t)$ is straightforward: in this case, $\bar{\rho} (t)$ is independent
of $t$ and the standard linear response theory approach to equilibrium systems yields

\begin{eqnarray}
\sigma (t<t_*) &=&-\frac{i }{{\cal N}} \sum_{\bf k}  \frac{1}{4(\epsilon_{\bf k}^{(0)})^2} \Biggl[ [j_{{\rm sp},x}^{(0)} ({\bf k}) ]_{-,+}  
[j_{{\rm sp},y}^{(0)} ({\bf k}) ]_{+,-}\nonumber \\ &-&   [j_{{\rm sp},y}^{(0)} ({\bf k}) ]_{-,+}  
[j_{{\rm sp},x}^{(0)} ({\bf k}) ]_{+,-} 
\Biggr]  
\:\:\:\: ,
\label{cch.w2}
\end{eqnarray}
\noindent
with the sum over ${\bf k}$ taken over the full Brillouin zone. Defining $\hat{d}_{\bf k}^{(0)}$ as 

\begin{eqnarray}
\hat{d}_{\bf k}^{z,(0)} &=& \frac{\xi_{\bf k}}{\epsilon_{\bf k}^{(0)}} \; , \nonumber \\
\hat{d}_{\bf k}^{x,(0)} &=& \frac{\Re e [-\Delta_{\bf k}^{ (0)} ]}{\epsilon_{\bf k}^{(0)}} \; , \nonumber \\
\hat{d}_{\bf k}^{y,(0)} &=& \frac{\Im m [\Delta_{\bf k}^{(0)} ]}{\epsilon_{\bf k}^{(0)}} \;\;\; , 
\label{extra.z1}
\end{eqnarray}
\noindent
and going through standard formal manipulations, we eventually obtain 

\begin{eqnarray}
&&\sigma {\color{black}(t<t_*)}=
\label{cz.5} \\
&& \frac{1}{16 \pi^2} \int_{\rm B.Z.} d^2k\: \left\{ \hat{d}^{(0)}_{\bf k} \cdot \left[ \frac{ \partial \hat{d}^{(0)}_{\bf k} }{\partial k_x} 
\times  \frac{ \partial \hat{d}^{(0)}_{\bf k} }{\partial k_y} \right] \right\}=\frac{ {\cal C}_0  }{2\pi} \nonumber 
\:\:\:\: . 
\end{eqnarray}
\noindent
In Eq.(\ref{cz.5}) we have denoted with ${\cal C}_0$ the first 
 Chern number of our specific model \cite{Qi2010}.
In general, our model Hamiltonian $\hat{H}_{\rm MF}$ is characterized by broken time-reversal symmetry but
 particle-hole symmetry \cite{Schnyder2009,Bernevig2013}. In particular, the latter
 symmetry is realized as 
 
\beq
\hat{H}_{\rm MF}^* (-{\bf k}) = U_c \hat{H}_{\rm MF} ({\bf k}) U_c^{-1}\;\; , 
\label{defC}
\eneq
 \noindent
with $U_c = \sigma_2$, $U_c^* U_c = - {\bf I}_{2 \times 2}$, $\sigma_2$ denoting the second Pauli matrix and ${\bf I}_{2 \times 2}$ 
the two-by-two identity. When  both $\Delta_{x^2-y^2}$ and $\Delta_{xy}$ are $\neq 0$, our model
can be obtained by continuously deforming the Hamiltonian in Eqs.(1,2) of Ref. \cite{Chern2016}, which is topologically
nontrivial. Therefore,   we infer that, in this case,  our system lies within a topologically superconducting 
phase, with Chern number ${\cal C}_0=\pm2$, depending on the relative sign of $\Delta_{x^2-y^2}^{(0)}$ and of 
$\Delta_{xy}^{(0)}$.  At variance, if $\Delta_{x^2-y^2}^{(0)} =0$, the system is within a topologically trivial case, and 
${\cal C}_0=0$, regardless of whether $\Delta_{xy}=0$  or not. 
  
To compute $\sigma (t)$ for $t>t_*$ it is operationally useful to split the integral in
Eq.(\ref{cch.9}) according to 

\begin{eqnarray}
&& \sigma (t)=  - i \lim_{\omega_0 \to 0 } \: \{ \int_{-\infty}^{t_*} + \int_{t_*}^t \}  \: d \tau \: \times \nonumber \\
&&   {\rm Tr} \{ [ j^{{\color{black} M}}_{{\rm sp} , x} (t) , 
j^{{\color{black} M}}_{{\rm sp} , y} ( \tau ) ] \rho (t ) \} \: \frac{ \sin ( \omega_0 ( \tau-t_*) )}{ \omega_0} e^{ \eta (\tau - t_*)}  \nonumber \\
&& \equiv \sigma^A(t)+\sigma^B(t)
\;\;\;\; .
\label{cch.z9}
\end{eqnarray}
\noindent
 In particular, for $\sigma^A (t)$ one gets 
 
 \begin{widetext}
 \begin{eqnarray}
 \sigma^A (t) &=& \frac{i}{{\color{black} {\cal N}}} \sum_{\bf k} \sum_{\lambda , \lambda'} \sum_{\mu , \nu} 
  \frac{1}{4 [ \epsilon_{\bf k}^{(0)}]^2}  e^{ i \epsilon_{\bf k}^{(1)} ( \lambda - \lambda') ( t-t_*)} 
 [j_{{\rm sp},x}^{(1)} ({\bf k} )]_{\lambda , \lambda'} 
 [ j_{{\rm sp},y}^{(0)} ({\bf k}) ]_{\mu,\bar{\mu}} \nonumber \\
 &\times& [ [ {\cal S}_{{\bf k};(\mu,\lambda')}^{(0,1)} ]^*  {\cal S}_{{\bf k};(\bar{\mu}, \nu  )}^{(0,1)} 
\bar{c}_{{\bf k};(\lambda,\nu)} (t) -  [ {\cal S}_{{\bf k};(\mu,\nu)}^{(0,1)} ]^*  {\cal S}_{{\bf k};(\bar{\mu},\lambda)}^{(0,1)} 
\bar{c}_{{\bf k};(\nu,\lambda')} (t) ] 
\;\;\;\; ,
\label{sigmaa.1}
\end{eqnarray}
\end{widetext}
\noindent
with $[j_{{\rm sp},y}^{(0)}]_{\mu,\mu'}$ being the matrix element of $j_{{\rm sp},y}^{(0)}$ in the 
basis of the eigenmodes  of $\hat{H}_{\rm MF}(t)$ for $t<t_*$ and  ${\cal S}_{{\bf k} }^{(0,1)} $ being the 
unitary transformation matrix defined by 

\beq
\Gamma_{{\bf k},\lambda}^{(0)} = \sum_{\mu} {\cal S}^{(0,1)}_{{\bf k};(\lambda,\mu)} \Gamma_{{\bf k},\mu}^{(1)} 
\:\:\:\: . 
\label{sigmaa.2}
\eneq
\noindent
By considering that  $\bar{c}_{{\bf k};(\alpha,\beta)} (t) \to_{t\to \infty} \delta_{\alpha,-}\delta_{\beta,-}$,
we readily obtain that the asymptotic form of $\sigma^A (t)$ for $2g(t-t_*) \gg 1$ is given by 

\begin{widetext}
\begin{eqnarray}
\sigma^A (t)&\to& \frac{i}{{\color{black} {\cal N}}} \sum_{\bf k}\sum_\mu   \frac{1}{4[\epsilon_{\bf k}^{(0)} ]^2} 
\Biggl[ e^{-2i\epsilon_{\bf k}^{(1)} (t-t_*) } [j_{{\rm sp},x}^{(1)} ({\bf k})]_{-,+} [j_{{\rm sp},y}^{(0)} ({\bf k})]_{\mu,\bar{\mu}} 
[ {\cal S}_{{\bf k};(\mu,+)}^{(0,1)}]^* S_{{\bf k};(\bar{\mu},-)}^{(0,1)} \nonumber \\
&-& e^{ 2i\epsilon_{\bf k}^{(1)} (t-t_*) } [j_{{\rm sp},x}^{(1)} ({\bf k})]_{+,-} [j_{{\rm sp},y}^{(0)} ({\bf k})]_{\mu,\bar{\mu}} 
[ {\cal S}_{{\bf k};(\mu,-)}^{(0,1)}]^* S_{{\bf k};(\bar{\mu},+)}^{(0,1)}    \Biggr]  
\:\:\:\: . 
\label{sigmaa.3}
\end{eqnarray}
\end{widetext}
\noindent
As $t-t_*$ gets large, on employing the stationary phase 
approximation similarly to what has been done in Ref.\cite{Derrico2020} for a 1D lattice model, 
and assuming that  $\epsilon_{{\bf k}}$ is fully gapped, as  in the case 
of a  d+id  order parameter, 
the right-hand side of Eq.(\ref{sigmaa.3}) 
can be readily shown to  decay like $(t-t_*)^{-1}$.  
 At variance, for $\sigma^B(t)$ we obtain
 
\begin{widetext}
\begin{eqnarray}
\sigma^B(t)&=&- \frac{i}{{\color{black} {\cal N}}}\sum_{\bf k}  \frac{1}{4 [\epsilon_{\bf k}^{(1)}]^2} \sum_\lambda \bar{c}_{{\bf k},(\lambda,\lambda)}(t) 
\Biggl[ (1-e^{2i\lambda \epsilon_{\bf k}^{(1)} (t-t_*)} )  [j_{{\rm sp},x}^{(1)} ({\bf k)}]_{\lambda,\bar{\lambda}}  
[j_{{\rm sp},y}^{(1)} ({\bf k})]_{\bar{\lambda},\lambda}
 \label{cch.z10} \\
&-& (1-e^{- 2i\lambda \epsilon_{\bf k}^{(1)} (t-t_*)} ) 
[j_{{\rm sp},y}^{(1)} ({\bf k})]_{\lambda,\bar{\lambda}}  [j_{{\rm sp},x}^{(1)} ({\bf k})]_{\bar{\lambda},\lambda} \Biggr]  \nonumber \\
 &+&  \frac{i}{{\color{black} {\cal N}}}\sum_{\bf k}   \frac{1}{4 [\epsilon_{\bf k}^{(1)}]^2} \sum_\lambda 
 (1-e^{2i\lambda \epsilon_{\bf k}^{(1)} (t-t_*)} ) \bar{c}_{{\bf k},(\lambda,\bar{\lambda})} (t) 
 \Biggl[  [j_{{\rm sp},x}^{(1)} ({\bf k})]_{\lambda,\lambda}  [j_{{\rm sp},y}^{(1)} ({\bf k})]_{\lambda,\bar{\lambda}} - 
 [j_{{\rm sp},y}^{(1)} ({\bf k})]_{\lambda, \bar{\lambda}}  [j_{{\rm sp},x}^{(1)} ({\bf k})]_{ \bar{\lambda}, \bar{\lambda}} \Biggr]    \nonumber
  \:\:\:\: . 
 \end{eqnarray}
 \end{widetext}
 \noindent
Apparently, for $2g(t-t_*) \gg 1$, we get

\begin{widetext}
\begin{eqnarray}
\sigma^B (t) &\to&   - \frac{i}{{\color{black} {\cal N}}}\sum_{\bf k}   \frac{1}{4 [\epsilon_{\bf k}^{(1)}]^2}  
\Biggl[ (1-e^{-2i  \epsilon_{\bf k}^{(1)} (t-t_*)} )  [j_{{\rm sp},x}^{(1)} ({\bf k})]_{-,+}  [j_{{\rm sp},y}^{(1)} ({\bf k})]_{+,-} \nonumber 
\\
&-& (1-e^{  2i  \epsilon_{\bf k}^{(1)} (t-t_*)} ) 
[j_{{\rm sp},y}^{(1)} ({\bf k})]_{-,+}  [j_{{\rm sp},x}^{(1)} ({\bf k})]_{+,-} \Biggr]   \equiv \sigma_I^B + \sigma^B_{II} (t) 
\label{cch.z12}  
 \:\:\:\: . 
 \end{eqnarray}
 \end{widetext}
 \noindent
  The former contribution,
$\sigma^B_I$, is independent of $t$ and is given by 

\begin{eqnarray}
\sigma^B_I &=& -\frac{i }{{\color{black} {\cal N}}} \sum_{\bf k}   \frac{1}{4(\epsilon_{\bf k}^{(1)})^2} \Biggl[ [j_{{\rm sp},x}^{(1)} ({\bf k}) ]_{-,+}  
[j_{{\rm sp},y}^{(1)} ({\bf k}) ]_{+,-}  \nonumber \\
&-&   [j_{{\rm sp},y}^{(1)} ({\bf k}) ]_{-,+}  
[j_{{\rm sp},x}^{(1)} ({\bf k}) ]_{+,-} 
\Biggr]  
\:\:\:\: ,
\label{cch.t2}
\end{eqnarray}
\noindent
which is the same form as the spin-Hall conductance for $t<t_*$ in Eq.(\ref{cch.w2}) and, accordingly, 
can be rewritten in the same form as in Eq.(\ref{cz.5}) as  

\begin{eqnarray}
&& \sigma {\color{black} (t\to\infty)} 
\label{cz.t5} \\
&& =\frac{1}{16 \pi^2} \int_{\rm B.Z.} d^2k\: \left\{ \hat{d}^{(1)}_{\bf k} \cdot \left[ \frac{ \partial \hat{d}^{(1)}_{\bf k} }{\partial k_x} 
\times  \frac{ \partial \hat{d}^{(1)}_{\bf k} }{\partial k_y} \right] \right\} =\frac{{\cal C}_1}{2\pi} \nonumber
\:\:\:\: , 
\end{eqnarray}
\noindent
with $\hat{d}^{a,(1)}_{\bf k}$ ($a=x,y,z$) defined as $\hat{d}^{a,(0)}_{\bf k}$ in Eq.(\ref{extra.z1}), 
with $0\to 1$, and ${\cal C}_1$ being the Chern number of the system in the $t\to\infty$ limit. 
In addition, there is the oscillating, time-dependent contribution which, again, employing the 
stationary phase approximation can be readily shown to decay
as $(t-t_*)^{-1}$,  just as $\sigma^A (t)$, for large values of $t-t_*$.

{\color{black} 
\subsection{Lindblad Master Equation approach to the nonequilibrium dynamics of an open,  two-dimensional 
superconducting system}
\label{smsec.3}

Our derivation of the time-dependent density matrix  of the nonequilibrium two-dimensional superconducting system coupled to the 
external bath is based on Eq.(\ref{leq.1}) for $\rho (t)$. Here, we ground our LME approach 
by sketching its derivation from a general model for a two-dimensional superconductor coupled to a metallic lead working as 
a dissipative bath \cite{Heimes2014}. Moreover, following the discussion of Refs.\cite{Yuzbashyan2005,Cui2019}, we evidence 
how one recovers our Eq.(\ref{leq.1}) form a completely 
different perspective, that is, within a  semiphenomenological formalism that accounts for the damping effects arising  from 
the superconductor dynamics  beyond BCS mean-field theory, such as interactions among the quasiparticles within the superconductor, or 
between the quasiparticles and the fluctuations of the superconducting order parameter \cite{Cui2019,longer_paper}. That the LME
in Eq.(\ref{leq.1}), pertinently complemented within the time-dependent SCMF approximation (which we
discuss above), accounts for two completely different realizations of the external bath, evidences the wide
applicability of our approach to solid-state systems such as the one we consider in our main paper. In addition, it is also worth stressing
that LME in many cases provides a good approximation to describe the dissipative dynamics in the case of cold atoms loaded over
optical lattices, coupled to a continuum of radiation modes \cite{Goldman2016}. Thus, it apparently applies equally well to open 
solid-state, as well as optical, systems. 

A minimal microscopic model for a two-dimensional superconductor coupled to a metallic lead can be recovered by properly adapting the
formalism of Ref.\cite{Heimes2014}. Prior to coupling the superconductor to the bath, we describe the former, within mean field 
approximation, by means of an Hamiltonian $\hat{H}_{\rm MF}$ in the same form as in Eq.(\ref{cch.1}) but, of course, without 
an explicit dependence on time $t$. At variance, denoting with  $d_{{\bf q},\sigma},d_{{\bf q},\sigma}^\dagger$ the single-fermion 
annihilation/creation operators in the normal lead, we describe it by means of the normal Hamiltonian $\hat{H}_N$, given by 

\beq
\hat{H}_N = \sum_{\bf q}\sum_\sigma \xi_{\bf q} d_{{\bf q},\sigma}^\dagger d_{{\bf q},\sigma}
\;\;\; , 
\label{men.1}
\eneq
\noindent
with $\xi_{\bf q}$ being the dispersion relations of quasiparticles in the normal contact. 
Finally, we encode the coupling between the superconductor and the contact in the 
single-particle tunneling Hamiltonian $\hat{H}_T$, given by 

\beq
\hat{H}_T=\sum_{{\bf k},{\bf q}}\sum_\sigma t_{{\bf k},{\bf q}} \{ c_{{\bf k},\sigma}^\dagger 
d_{{\bf q},\sigma} + d_{{\bf q},\sigma}^\dagger c_{{\bf k},\sigma} \} 
\:\: . 
\label{men.2}
\eneq
\noindent
Regarding  $\hat{H}_T$ in Eq.(\ref{men.2}) we point out that, although it is linear in the single-fermion operators of the superconductor
and of the lead, when going along the derivation of the LME, 
composite operators of the basic jump operators can also account for multi-particle dissipative
processes, such as two-particle losses \cite{Kantian2009,Mazza2023}. 
Yet, those processes  enter to higher order in the expansion in the coupling 
between the superconducting system and the bath and, therefore, being subleading with respect to the single quasiparticle 
exchange,  we do not explicitly consider them here.  

To derive the LME for the superconductor density matrix operator $\rho (t)$, we begin with considering the time evolution 
operator for the density matrix operator of the whole system, $\rho_w (t)$, to second order in $\hat{H}_T$. Assuming that 
$\hat{H}_T$ is turned on at $t=0$, this reads 
\cite{Petruccione2002} 

\beq
\frac{d \rho_w(t)}{dt}= -\int_0^t\: dt'\: [\hat{H}_T(t),[\hat{H}_T(t'),\rho_w(t)]]
\;,
\label{men.3}
\eneq
\noindent
with the operator $\hat{H}_T (t)$ and $\rho_w (t)$ taken in the interaction representation corresponding to the decoupled system. 
Within weak coupling hypothesis between the system and the lead, we assume that, consistently 
with the standard Markov approximation,  the bath stays at 
equilibrium at any time $t$. We therefore set $\rho_w (t)=\rho (t)\otimes \rho_B$, with the bath density
matrix $\rho_B=e^{-\beta \hat{H}_N}/{\rm Tr} [e^{-\beta \hat{H}_N}]$. Tracing over the bath degrees of freedom and 
assuming translational invariance in real space, which implies $t_{{\bf k},{\bf q}}=t_{{\bf k}} \delta_{{\bf k},{\bf q}}$, 
Eq.(\ref{men.3}) yields, after integrating over $dt'$ and resorting to the Schr\"odinger representation

\begin{eqnarray}
&& \frac{d \rho (t)}{dt} = -i[H_{\rm MF},\rho(t)]+ \sum_{\lambda = \pm}\sum_{\bf k}  T_{\bf k} \label{men.4} \\
&&[ f(-\lambda \epsilon_{\bf k} ) (
 2 \Gamma_{{\bf k},\lambda}   \rho (t) \Gamma_{{\bf k},\lambda}^\dagger  -
\{ \Gamma_{{\bf k},\lambda}^\dagger  \Gamma_{{\bf k},\lambda} ,\rho (t)  \}  ) 
  \nonumber \\
&&+ f(\lambda \epsilon_{\bf k} (t) ) ( 2 \Gamma_{{\bf k},\lambda}^\dagger \rho (t) 
\Gamma_{{\bf k},\lambda} - \{  \Gamma_{{\bf k},\lambda} \Gamma_{{\bf k},\lambda}^\dagger , \rho (t) \} ) ] \nonumber 
\:  .
\end{eqnarray}
\noindent 
with $T_{\bf k}=2\pi t_{\bf k}^2$ and the quasiparticle operators and the other parameters defined according to Eqs.(\ref{cch.z4},\ref{cch.z4a})
above. 

On neglecting, just for the sake of simplicity, the dependence of $T_{\bf k}$ on ${\bf k}$ and setting $T_{\bf k}=g$ and on 
employing the time-dependent self-consistent approximation described above, we just recover Eq.(\ref{leq.1}).

A different route also leading to Eq.(\ref{leq.1})  goes through phenomenologically accounting 
for damping effects that go beyond mean-field BCS approaximation. Specifically, one begins with the Bloch-like equation for 
the Nambu pseudospin at a given ${\bf k}$, ${\bf S}_{\bf k}$, defined as 

\beq
{\bf S}_{\bf k} = \frac{1}{2} [c_{{\bf k},\uparrow}^\dagger , c_{{\bf -k},\downarrow}]\vec{\sigma} \left[\begin{array}{c}
c_{{\bf k},\uparrow} \\ c_{{\bf -k},\downarrow}^\dagger \end{array} \right]
\:\:\:\:.
\label{men.5}
\eneq
\noindent
For a system described by $\hat{H}_{\rm MF}(t)$  in Eq.(\ref{cch.1}), the equation of motion for the time dependent 
average, $\langle {\bf S}_{\bf k} (t) \rangle$, is given by \cite{Cui2019,longer_paper}

\beq
\frac{d \langle {\bf S}_{\bf k} (t)\rangle}{dt} = {\bf B}_{\bf k} (t) \times \langle {\bf S}_{\bf k} (t)\rangle 
\;\; , 
\label{men.6}
\eneq
\noindent
with ${\bf B}_{\bf k} (t) = [ -\Re e [\Delta_{\bf k} (t)],\Im m [\Delta_{\bf k}(t)],\xi_{\bf k}]^T$. In order to phenomenologically
account for damping effects, in analogy to the spin precession problem, 
one adds to the right-hand side of Eq.(\ref{men.6}) the typical terms that encode the longitudinal ($T_1^{-1}$) and the 
transverse ($T_2^{-1}$) relaxation rates. On setting $T_1^{-1}=T_2^{-1}=2g$, Eq.(\ref{men.6}) is modified into

\beq
\frac{d \langle {\bf S}_{\bf k} (t)\rangle}{dt} = {\bf B}_{\bf k} (t) \times \langle {\bf S}_{\bf k} (t)\rangle -2g \langle {\bf S}_{\bf k} (t)\rangle
+ 2g\langle {\bf S}_{{\bf k},*}(t)\rangle 
\;\; , 
\label{men.7}
\eneq
\noindent
with $\langle {\bf S}_{{\bf k},*}(t)\rangle $ being the thermalized spin configuration at time $t$. Eq.(\ref{men.7}) is 
exactly recovered within our LME framework by noting that one can express $\langle {\bf S}_{\bf k}(t)\rangle$ 
as 

\beq
\langle {\bf S}_{\bf k} (t)\rangle = \left[\begin{array}{c} f_{\bf k}^* (t) \\ f_{\bf k} (t) \\ \nu_{\bf k} (t) \end{array}\right] 
\;\;\;\; . 
\label{men.8}
\eneq
\noindent
Given Eq.(\ref{men.8}), we use Eqs.(\ref{s.0.2}) to write the equation of motion for $\langle {\bf S}_{\bf k}(t)\rangle$.

It is now easy to check that one exactly recovers Eq.(\ref{men.7}), with $\langle {\bf S}_{\bf k}(t)\rangle =  {\bf B}_{\bf k}(t)/2\epsilon_{\bf k}(t)$. 
We therefore conclude that, as stated above, also the phenomenological approach of Ref.\cite{Cui2019} gives us back 
Eq.(\ref{leq.1}).
}

\subsection{Rate function and spin-Hall conductance across the d+id$\to$s topological phase transition}
\label{smsec.4}

In the main text, we analyze the id$\to$d+id DPT as an explicit example of a TDPT between a topologically trivial and a 
topologically nontrivial state. As  an additional example, we now discuss the d+id$\to$s DPT as an alternative example of a 
TDPT between a topologically nontrivial and a topologically trivial phase. 
 Specifically, {\color{black} we choose a pure s-wave phase as our topologically trivial one. To trigger a TDPT toward 
 the s phase, at $t=0$ we turn on an additional, onsite attractive interaction in the spin singlet channel, with strength 
 equal to $U^{(1)}$.  In general, for $U,V$ and $Z$ all different from 0, Eqs.(\ref{eh.bis1})  generalizes to 
 \cite{longer_paper}
 
 \begin{eqnarray}
 \Delta_{\bf k} &=& \Delta_S + 2\Delta_{x^2-y^2} \{ \cos(k_x)-\cos (k_y) \} \nonumber \\
 &-&4i\Delta_{xy} \sin (k_x)\sin(k_y)
 \:\:\:\: ,
 \label{exsup.1}
 \end{eqnarray}
 \noindent
 with $\Delta_S$ determined by the self-consistent equation
 
 \beq
 \Delta_S = \frac{U}{2{\cal N}}\sum_{\bf k} \frac{\Delta_{\bf k}}{\epsilon_{\bf k}}
 \:\:\: . 
 \label{exsup.2}
 \eneq
 \noindent
 To define the nonequilibrium protocol corresponding to the TDPT described above, we 
  prepare our system in a d+id superconducting state with  $\Delta_S^{(0)}=0$, $\Delta_{x^2-y^2}^{(0)}=0.077$,}
and $\Delta_{xy}^{(0)} = 0.013$.  For $t>0$, we let the system evolve with the Hamiltonian $H_{\rm MF}$ with 
$Z^{(1)}=V^{(1)}=1.5$ and  $U^{(1)}=3.0$ and with the system-bath coupling $g=0.2$. 
To evidence the DPT, we first of all employ the approach we discuss in the main text, to compute $\Delta_S(t),\Delta_{x^2-y^2}(t)$,
and $\Delta_{xy}(t)$. In Fig.\ref{sdid_f} we plot the three time-dependent gaps as functions of $t$.  
We see that apparently   the system keeps within 
the d+id phase, with both $\Delta_{x^2-y^2}(t)$ and $\Delta_{xy}(t)$ finite and with 
$\Delta_S(t)=0$,  as long as $t<t_*$, with $t_*\approx 50$. At $t=t_*$, $\Delta_S (t)$ jumps to a finite value, goes 
through a small oscillating transient and eventually stabilizes over its constant, asymptotic value $\Delta_{S,\infty}=0.85$. 
At the same time, after a similar transient, $\Delta_{x^2-y^2}(t)$ and $\Delta_{xy}(t)$ take their asymptotic values,
$\Delta_{x^2-y^2,\infty}=\Delta_{xy,\infty}=0$. The phase with $\Delta_{S,\infty} \neq 0$ and $\Delta_{x^2-y^2,\infty}=
\Delta_{xy,\infty}=0$ is  topologically trivial. This is, in fact, the first piece of evidence that,  
at $t=t_*$, our system goes through a TDPT.  
    \begin{figure}
 \center
\includegraphics*[width=.9 \linewidth]{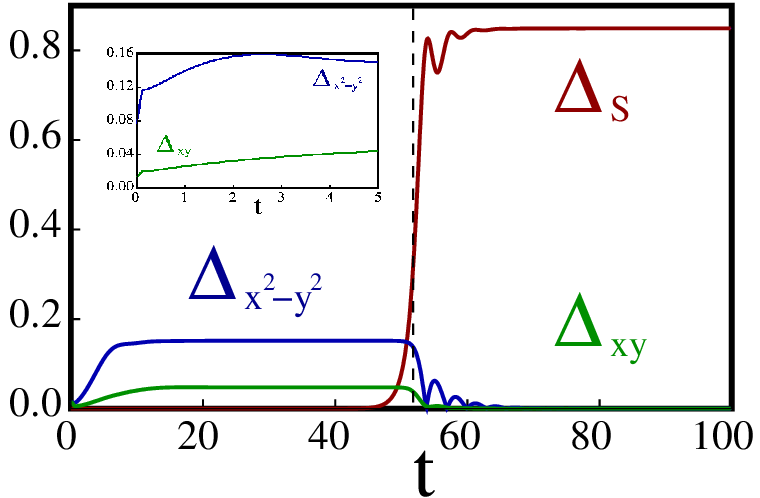}
\caption{{\color{black} Time dependent gap  $\Delta_S (t)$ (red curve), $\Delta_{x^2-y^2}(t)$ (blue curve),
and  $\Delta_{xy} (t)$ (green curve)}, computed 
in a system with   $\mu=0$, $U^{(1)}=30$ and $V^{(1)}=Z^{(1)}=1.5$,  with
 $g=0.2$. For $t<0$ the system is assumed to be prepared in a state with  $\Delta_S^{(0)}=0$, $\Delta_{x^2-y^2}^{(0)}=0.077$,
and $\Delta_{xy}^{(0)} = 0.013$. The dashed vertical line marks the DPT at $t=t_*\approx 50$. [{\bf Inset:} The same 
plot (for $\Delta_{x^2-y^2}(t)$ and $\Delta_{xy}(t)$) restricted to $0\leq t<5$].}
\label{sdid_f}
\end{figure}
 
As we have done in addressing the id$\to$d+id TDPT in the main text, we proceed by
 studying the rate function $\omega (t)$ for the d+id$\to$s PT. Computing $\omega (t)$ 
 with the procedure detailed in Ref.\cite{longer_paper}, we derive the plot of   Fig.\ref{echol}, 
 where we draw   $\omega (t)$ as a function of $t$. 
 We clearly see the nonanalyticity at the DPT, corresponding to a 
change in the slope of $\omega (t)$ at $t=t_*$, and an over-all increase of 
the rate function  at larger values of $t$, associated to the expected reduction of ${\cal F}(t)$ for $t>t_*$.

    \begin{figure}
 \center
\includegraphics*[width=.9 \linewidth]{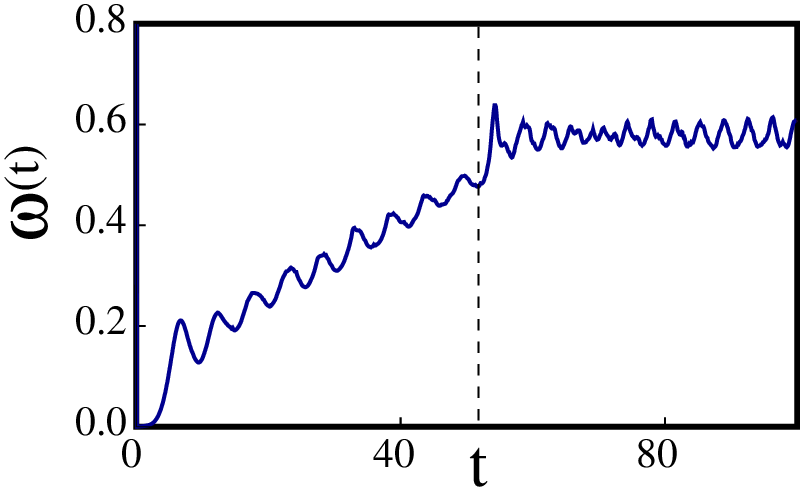}
\caption{   $\omega (t)$ as a function of $t$   with the time-dependent 
MF Hamiltonian with parameters $\Delta_{x^2-y^2} (t)$,  $\Delta_S (t)$ and $\Delta_{xy} (t)$, as in 
Fig.\ref{sdid_f}, computed for $g=0.2$.   
The dashed  vertical line  marks the DFT.}
\label{echol}
\end{figure}
\noindent

Finally, we now compute  $\sigma (t)$ across the DPT.  Again, we note that, since around 
$t=t_*$ the superimposed oscillations in  $\Delta_{S} (t), \Delta_{x^2-y^2}(t)$ and $\Delta_{xy}(t)$ 
have no substantial qualitative effects, the behavior of the superconducting gaps
can be well approximated {\color{black} by a sudden jump in their values at} $t=t_*$.
Accordingly, we   set    $\Delta_{\bf k} (t)=\theta (t_*-t) \Delta_{\bf k}^{(<)}  + \theta (t -t_*)\Delta_{\bf k}^{(>)}$, with 

\begin{eqnarray}
\Delta_{\bf k}^{(<)}&=&2\Delta_{x^2-y^2}^{(<)} \{\cos (k_x)-\cos(k_y)\}\nonumber \\
&-&4i\Delta_{xy}^{(<)} \sin(k_x) \sin (k_y) \;\; , \nonumber \\
\Delta_{\bf k}^{(>)} &=& \Delta_S^{(>)} 
\;\; , 
\label{total.1}
\end{eqnarray}
\noindent
and all the other system parameters chosen consistently with the numerical values of the parameters
corresponding to  Fig.\ref{sdid_f}, that is,  $\mu=0$, $g=0.2$, $\Delta_{x^2-y^2}^{(<)}=0.152$,
$\Delta_{xy}^{(<)} =0.048$, and $\Delta_{S}^{(>)}=0.849$.  After the {\color{black} sudden change in the 
superconducting gaps}, 
$\Delta_{\bf k} (t)$ is purely s-wave and, therefore, the corresponding superconducting phase is topologically 
trivial. Thus,  we expect that $\sigma(t)\to0$ as $t\to\infty$. 
Therefore, we conclude that  the time evolution reported in Fig.\ref{sdid_f} corresponds to a TDPT.  

 In Fig.\ref{total_sigma}  we plot our result for $\sigma (t)$, computed as discussed above.   
   With a dashed  horizontal line we highlight the asymptotic value 
  of $\sigma (t)$. As expected, we see that,  for $t-t_* < 0$, $\sigma (t)$ is constantly equal to 2 (in the units of the plot), as it must be, in 
  the stationary  topological phase, with both $\Delta_{x^2-y^2}$ and $\Delta_{xy}$ being $\neq 0$ \cite{Chern2016,longer_paper}. 
  At $t-t_* = 0$ (which, in Fig.\ref{total_sigma},
  we mark with a dashed  vertical line), $\sigma (t)$ shows a sudden jump to lower values and,  
  for $t-t_* > 0$, it starts to oscillate around the asymptotic value $\sigma^{(1)}= \sigma (t \to \infty ) = 0$.
This is appropriate for the  topologically trivial phase that asymptotically sets in, 
   with  $\Delta_S (t\to\infty) \neq 0$  and $\Delta_{x^2-y^2}(t\to\infty)=\Delta_{xy}(t\to\infty)=0$.
On increasing $t-t_*$, the oscillations of $\sigma (t)$ around $\sigma_\infty $ are apparently suppressed and
 $\sigma (t)$ flows toward its asymptotic value $\sigma^{(1)}  = 0$.

    \begin{figure}
 \center
\includegraphics*[width=.9 \linewidth]{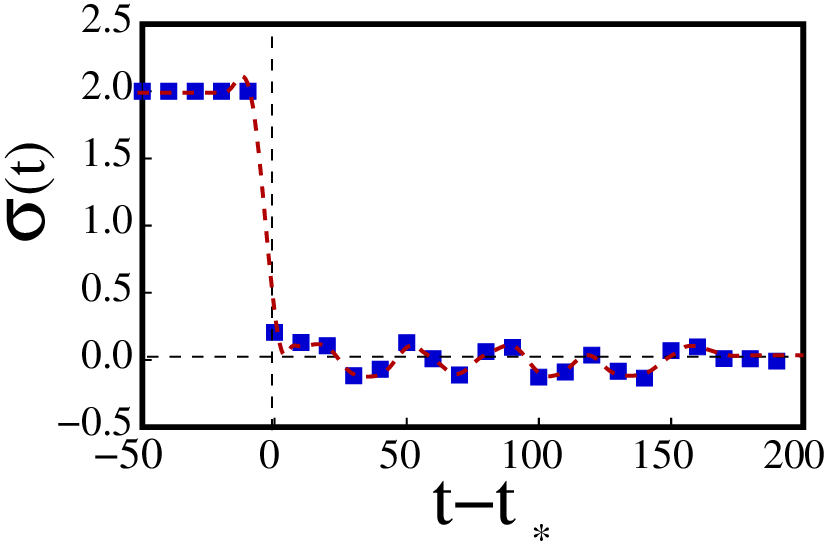}
\caption{Spin-Hall conductance
$\sigma (t)$ (in units of $(2\pi)^{-1}$)  computed  with the parameters set as 
specified in Eq.(\ref{total.1}) and in the related discussion.  For $t=t_*$ we have used the approximated value $t_* = 50$.
  The results are shown as a function of 
 $t-t_* \geq 0$ for $-50 \leq t-t_* \leq 200$. The dashed  horizontal line marks the asymptotic value 
 $\sigma (t \to \infty)$, the dashed   vertical line marks the DPT at $t-t_*=0$, and the dashed line  connecting the 
 dots is a guide to the eye only. }
\label{total_sigma}
\end{figure}
\noindent
Therefore, we   conclude that  the time evolution of the  superconducting gaps  
in Fig.\ref{sdid_f} does, in fact, correspond to a TDPT between a topologically nontrivial and a 
topologically trivial case. Aside for being relevant per se, our result has also potentially important
practical consequences. Indeed, we infer that, although, for $t>0$,  the system evolves with 
an Hamiltonian whose parameters, at equilibrium, would not correspond to a topological phase,
having prepared the system at $t=0$ in the groundstate of a topologically nontrivial Hamiltonian, 
the system keeps showing topological   properties at finite $t>0$ until it unavoidably 
switches to a trivial phase, at $t=t_*$. By analogy with the plots of Fig.\ref{sdid_f}, we infer that
the time of persistence of the topological phase ($t_*$) is determined by the strength of the system-bath coupling
 ($g$). Thus, $t_*$ should be, in principle, tunable, by pertinently
engineering the system parameters, which, for practical purposes, would allow the persistence, for a long 
 time, of a topological phase in a system whose dynamics is described by a nontopological
Hamiltonian. In such an approach, there is  no need of tuning and stabilizing the Hamiltonian parameters, but only of pertinently
setting the initial state of the system.   }

   \begin{acknowledgements}
We thank N. Lo Gullo and F. Plastina for insightful discussions. \\  A.N., C.A.P., L.L., and D.G.  
  acknowledge   financial support  from Italy's MIUR  PRIN project  TOP-SPIN  (Grant No. PRIN 20177SL7HC).  \\
 L.L. acknowledges financial support by a project funded under the National Recovery 
 and Resilience Plan (NRRP), Mission 4 Component 2 Investment 1.3 - Call for tender No. 341
  of 15/03/2022 of Italian Ministry of University and Research funded by the European Union – NextGenerationEU, award number PE0000023, Concession Decree No. 1564 of 11/10/2022 adopted by the Italian Ministry of University and Research, CUP D93C22000940001, Project title "National Quantum Science and Technology Institute" (NQSTI).\\
 A.N. and R.E. acknowledge  funding by the Deutsche Forschungsgemeinschaft (DFG, German Research Foundation) under Grant 
 No.~ 277101999, TRR 183 (project C01), under Germany's Excellence Strategy - Cluster of Excellence Matter
 and Light for Quantum Computing (ML4Q) EXC 2004/1 - 390534769, and under Grant No.~EG 96/13-1.        
 \end{acknowledgements}

\bibliography{lindblad_super}

\end{document}